\documentstyle[12pt,epsfig]{article}

\begin{document}
\begin{Large}
\begin{center}
{\bf A Strong Electroweak Sector \\ with Decoupling at Low Energy
\footnote{
Talk presented at the conference Beyond the Standard Model V, 
Balholm, Norway, April 29 - May 4, 1997.
This work has been carried out within the Program Human Capital and
Mobility: ``Tests of electroweak symmetry breaking
and future European colliders'', CHRXCT94/0579.}}
\end{center}
\end{Large}
\vskip 1.0cm
\centerline {D. Dominici}
\vskip 1cm

\noindent
\centerline{{\it Dipartimento di Fisica Universit\`a  di Firenze}}
\centerline{{\it I.N.F.N. Sezione di Firenze}}
\centerline{{\it Lgo E.Fermi 2, 50125 Firenze, Italia.}}
\hfill\break\noindent
\vskip 1cm

\begin{abstract}
I discuss the possible symmetries of effective theories for
spinless and spin one bosons, to
concentrate on a model with vector and axial vector strong
interacting bosons possessing a discrete symmetry demanding degeneracy
(degenerate BESS model,
BESS standing for Breaking 
Electroweak Symmetry Strongly). The phenomenology at future hadron
colliders
is also presented.
\end{abstract}

\section*{Introduction}
%\[
%\widehat{a} + \widehat{ab} + \widehat{abc} + \widehat{abcd}
%\]
%
%%\show\frak
% 
%\[
%%      {\bf x}^{\bf x} \triangleq z 
%      {\bf x}^{\bf x}\triangleq{z} \tensor{T} \frak{E^E}=\frak{mc}^2
%%      {\bf x}^{\bf x}\triangleq {z} \tensor{T} \frak{E}=\frak{mc}^2
%\]
% 
%\[
%{\Bbb {NQRZ}} \qquad \because \eth\ggg\bigstar \therefore\blacktriangleright\rightsquigarrow \blacksquare
%\]
% 
The standard model (SM) of the electroweak interactions is confirmed with
excellent
accuracy by the existing results at LEP, SLC and Tevatron. Therefore
only extensions which smoothly modify this theory are still conceivable.
The minimal supersymmetric standard model (MSSM) is the most favorite.
It solves the so-called hierarchy problem which is related to the
presence of quadratic mass divergences due to scalar fields 
and it has an interesting property of decoupling. In the limit in
which all
the supersymmetric particles become heavy one recovers the SM with a
light Higgs.

This decoupling property however is not peculiar of the MSSM only. In
fact in my talk I will present a model for a strong electroweak sector
sharing this property.

 Models of dynamical breaking of the electroweak
symmetry are based just on the symmetry structure, not taking into
account
 the underlying theory which is responsible of the breaking and
the corresponding chiral lagrangians are built using, as 
degrees of freedom, the
goldstone bosons, the longitudinal components of $W$ and $Z$.
Experiments, like the measurement of
the $\rho$ parameter,
 suggests the existence of a custodial symmetry $SU(2)$ \cite{cust}, which
guarantees $\rho=1$.
Therefore chiral lagrangians can be built 
as  a non linear $\sigma$-model assuming the spontaneous breaking 
  $SU(2)_L\otimes SU(2)_R \to
SU(2)_{L+R}$.

Effective lagrangians can be derived using the global symmetry 
$G=SU(2)_L\times SU(2)_R$ and an expansion in the energy \cite{wphy}.
Goldstones are described by a unitary field $U(x)=\exp(i\pi^a(x)\tau^a/v)$,
 transforming  under $G$
as $(2,2)$ or $U\to g_L Ug_R^\dagger$. 
The most general chiral lagrangian is a sum of infinite number of
terms
with an increasing number of derivatives or equivalently in terms
of increasing powers of
energy or momentum. 
These lagrangians allow for a description of the physics below a
cutoff scale $\Lambda$, related to
the new physics.
Vector resonances can be introduced in chiral lagrangians by following
Weinberg \cite{weinberg} or in equivalent way by means of the hidden
 local symmetry
approach \cite{bando}. 
The model so obtained when this technique is applied to the 
electroweak symmetry breaking sector is called BESS \cite{bess}.

 The model we will review here is a model with
new vector and axial vector 
bosons degenerate in mass. 
These bosons correspond to the gauge bosons associated to a hidden symmetry,
$H'=SU(2)_L \otimes SU(2)_R$. The symmetry group of the theory becomes
$G'=G\otimes H'$. It breaks down spontaneously to $H_D=SU(2)$,
 the diagonal subgroup of $G^\prime$, and gives rise to
nine Goldstones. Six of these are absorbed by the vector and
axial vector bosons. As soon as we perform
the gauging of the subgroup $SU(2)_L \otimes U(1)_Y \subset G$, 
the three remaining Goldstones disappear giving masses to the 
SM gauge bosons. This general procedure
for building models with vector and axial vector
resonances  is discussed in \cite{assiali}, while this special model
has been proposed in 
\cite{debess}. This model has the nice property   that all the
deviations in the low 
energy parameters from their SM values are strongly 
suppressed. This allows the existence of a strong electroweak
sector
 at 
relatively low energies within the precision of electroweak tests, 
such that it may be accessible with accelerators designed for the near future.
As such it offers 
possibilities of experimental tests even with future or existing
machines of relatively low energy. In the following 
 phenomenological implications at Tevatron upgrade and LHC will be 
discussed \cite{degelhc}.

\section*{Degenerate BESS Model}

The model includes two new triplets of vector bosons ($L^\pm$,$L_3$)
and ($R^\pm$,$R_3$). The parameters of 
the model are a new gauge coupling constant $g''$ and a mass parameter $M$, 
related to the scale of the underlying symmetry breaking sector.
 In the
following we give approximate formulas in the limit $M \to \infty$ and
$g'' \to \infty$. For the numerical analysis the exact
formulas of \cite{debess} were used.

In the charged sector the fields $R^\pm$ are unmixed for any value of ${g''}$.
Their mass is given by:
\begin{equation}
M^2_{R^\pm} \equiv M^2
\label{8.2}
\end{equation}

The charged fields $W^\pm$ and $L^\pm$ have the following masses:
\begin{equation}
M^2_{{W}^\pm}=\frac{v^2}{4} { g}^2
, ~~~
M^2_{{L}^\pm}=M^2 (1+2 {x^2})
\label{8.5}
\end{equation}
where $x=g/g''$,  $g$ is the usual $SU(2)$
gauge coupling constant and $v^2=1/({\sqrt{2} G_F})$.

In the neutral sector we have:
\begin{equation}
M^2_{Z}=\frac{M^2_W}{c^2_{\theta}} ~~~
M^2_{L_3}=M^2\left(1+2 x^2\right) ~~~
M^2_{R_3}=M^2\left(1+2 x^2 \tan^2 \theta\right)
\label{8.16}
\end{equation}
where $\tan \theta = s_{\theta}/c_{\theta} = g'/g$ and $g'$ is the usual 
$U(1)_Y$ gauge 
coupling constant. Notice that for small $x$ all the new vector
resonances are degenerate in mass.

The charged part of the fermionic lagrangian is
\begin{equation}
{\cal L}_{charged}=
-\left(a_W W_\mu^-+a_L L_\mu^-\right)J_L^{(+)\mu}+ H.c.
\end{equation}
where
\begin{equation}
a_W=\frac {g}{\sqrt{2}} ~~
a_L= -g x
\label{9.2}
\end{equation}
apart from higher order terms and $J_L^{(+)\mu}={\bar \psi}_L \gamma^{\mu}
\tau^+ \psi_L$ with $\tau^+=(\tau_1 + i\tau_2)/2$.
 Let us notice that the $R^{\pm}$ are not coupled 
to the fermions.

\begin{figure} % fig 1
\vspace{10pt}
\centerline{\epsfig{file=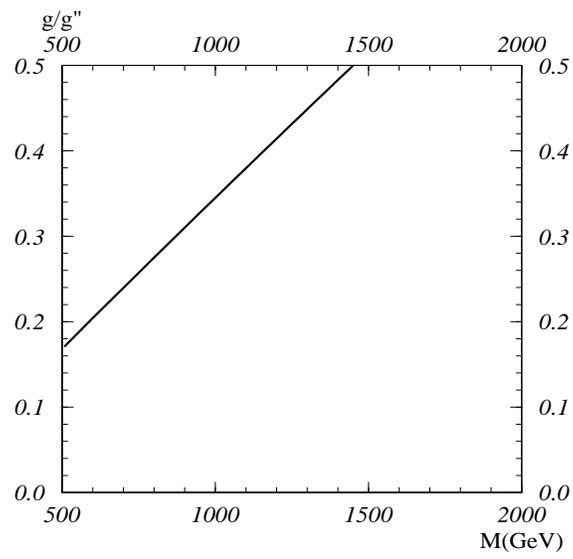,height=3.0in,width=3.5in}}
\vspace{10pt}
\caption{$ 90\% $ C.L. upper bounds on $g/g''$ vs. $M$ from 
LEP/Tevatron/SLC data.}
\label{fig1}
\end{figure}

For the neutral part we get
\begin{eqnarray}
{\cal L}_{neutral}&=& -\Big\{ eJ_{em}^\mu\gamma_\mu +
\left[A J_L^{(3)\mu}+BJ_{em}^\mu \right]Z_\mu\nonumber\\
&+&\left[C J_L^{(3)\mu}+DJ_{em}^\mu \right]L_{3\mu}\nonumber\\
&+&\left[E J_L^{(3)\mu}+FJ_{em}^\mu \right]R_{3\mu}\Big\}
\end{eqnarray}
where $\gamma_\mu$ in the preceeding formula is the photon field and 
again in the limit $M\to\infty$, $x\to 0$,
\begin{eqnarray}
&A&= \frac{g}{c_{\theta}}
~~~~B= -\frac{g s^2_\theta}{c_\theta} \nonumber \\
&C&=-\sqrt{2} g x ~~~~D= 0\nonumber \\
&E&= \sqrt{2} g \frac{x}{c_\theta} \tan^2 \theta ~~~~F= -E
\end{eqnarray}
and $J_{em}^{\mu}= Q {\bar \psi} \gamma^{\mu} \psi$, 
$J_L^{(3)\mu}={\bar \psi}_L \gamma^{\mu} T_L^3 \psi_L$ are the usual neutral
currents.

In Fig. \ref{fig1}
 $ 90\% $ C.L. upper bounds on $g/g''$ versus  $M$ from 
LEP, Tevatron and SLC data are shown. 
The  limits  (continuous line) are obtained calculating virtual
effects up to order $M_W^2/M^2$, and using the experimental data from ref.
\cite{altarelli}. Note that in the low energy limit 
($M \to \infty$) there are no deviations from the SM, thus 
allowing to consider light new resonances for the strong sector.

\section*{Degenerate BESS at Tevatron and LHC}

 We will now study the detection of charged and neutral vector resonances 
from a strong electroweak sector at the upgrading of the Fermilab
Tevatron~\cite{TeV2000}. The option we have chosen is the so called 
TeV-33, with 
a c.m. energy of the collider of 2 TeV and an integrated luminosity of 
$10~fb^{-1}$. 
We have considered the total 
 cross-section $p {\bar p} \to L^{\pm},W^{\pm}\to
\mu \nu_\mu$ and compared it with the SM background.

\begin{table}
\hfil
\vbox{\offinterlineskip
\halign{&#& \strut\quad#\hfil\quad\cr
\hline
\hline
&$g/g''$ && $M$ && $\Gamma_{L^\pm}$ && $S/{\sqrt{S+B}}$ & \cr
\hline
&\omit&&GeV&&GeV&&\omit& \cr
\hline
&\omit&&\omit&&\omit&&\omit&\cr
&0.12 && 400 && 0.4 && 24.9 & \cr  
&0.20 && 600 && 1.7 && 15.4 & \cr  
&0.40 && 1000 && 11.1 && 4.0 & \cr 
&\omit&&\omit&&\omit&&\omit&\cr
\hline
\hline}}
\caption{Degenerate BESS at TEV-33 for the process
 $p\bar p\to L^\pm\to
\mu\nu_\mu+X$.}
\end{table}

Up to a region of $1~ TeV$ the limits from TeV-33
option are stronger with respect
to LEPI (we recall that LEPII will only marginally improve LEPI
results \cite{debess}).

The events where simulated using Pythia Montecarlo \cite{phy}. The 
simulation was performed using the expected detector resolution, in particular
a smearing of the energy of the leptons was done according to
$\Delta E/{\sqrt{E}} = 10\%$ and the error on  the 3-momentum determination 
was assumed to be between $3\%$ for a mass of the order of $500~ GeV$
 and $5\%$ for 
a mass of $1000~ GeV$. 

We have examined various cases with different choices of $M$ and $g/g''$
(taken inside the physical region shown in Fig. 1)
to give an estimate of the sensitivity of the model to this
option for the upgrading  of the Tevatron (see Table 1
for
the charged case). 

For each case we have selected cuts to maximize the statistical 
significance of the signal.
We see that the number of signal events decreases for increasing mass of the 
resonance. The conclusion is that Tevatron with the high luminosity option 
will be able to discover a strong electroweak resonant sector as 
described by the degenerate BESS model for masses up to $1~ TeV$. It can be 
seen from the calculation of the statistical significance (see \cite{degelhc})
that the charged process allows to push further the discovery limits
of the new vector bosons  with respect to the neutral
process.
However the experimental check of the model requires the proof of the
existence of both neutral and charged vector bosons.
Notice that the reconstruction of the resonance mass 
requires a careful study 
of the experimental setup, due to the smallness of the resonance width.

We have also considered the phenomenology
at LHC with a center of mass  energy $\sqrt{s}=14~ TeV$,
a luminosity of $10^{34} cm^{-2} s^{-1}$  and one year run ($10^7~s$) 
\cite{degelhc}. 

\begin{figure} % fig 1
\centerline{\epsfig{file=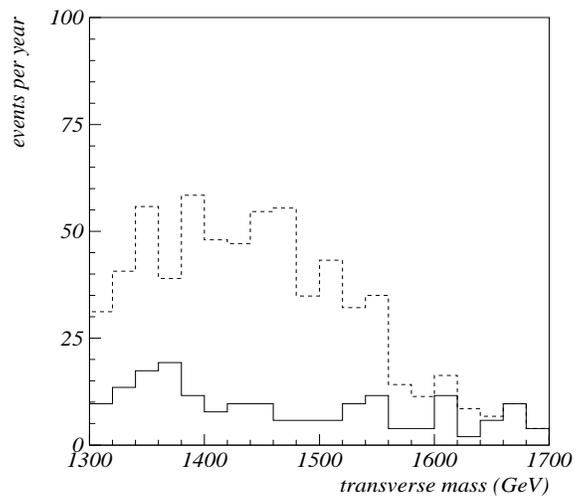,height=3.0in,width=3.5in}}
\vspace{10pt}
\caption{Transverse mass differential distribution of
$pp \to L^\pm, W^\pm \to \mu \nu_\mu$ events at LHC
 for $M=1500~ GeV$, 
$g/g''=0.1$.}
\label{fig2}
\end{figure}

In Fig.  \ref{fig2}
we show the differential distribution of events at LHC of 
$pp \to L^\pm , W^\pm \to \mu \nu_\mu$  in the transverse mass of the new vector
boson for $M=1500~GeV$ and $g/g''=0.1$. 
The following cuts have been applied: $\vert p_{T\mu}\vert >500~ GeV$,
$ m_{T} >1300 ~ GeV$. The number of signal events per year is 469, the 
corresponding background consists of 247 events.
 The energy of the 
muons was smeared by $10\%$ and the error in the 3-momentum increases with the 
momentum of the muon from $3\%$ to $9\%$ as stated in \cite{atlas}.

The statistical significance for some choices of
the parameters   is given in Table 2,
showing that the discovery of a charged resonance up to $2~ TeV$ with $g/g''
\simeq 0.1$ is well within the reach of LHC. The limit of detection for a 
$2~ TeV$ mass is reached for a value of $g/g''=0.03$.

I would like to thank R.Casalbuoni, P.Chiappetta,
A.Deandrea, S.De Curtis, F.Feruglio,
R.Gatto and M.Grazzini for the fruitful and enjoyable collaboration 
on the topics covered here.

\begin{table}
\hfil
\vbox{\offinterlineskip
\halign{&#& \strut\quad#\hfil\quad\cr
\hline
\hline
&$g/g''$ && $M$ && $\Gamma_{L^\pm}$ && $S/{\sqrt{S+B}}$ & \cr
\hline
&\omit&&GeV&&GeV&&\omit& \cr
\hline
&\omit&&\omit&&\omit&&\omit&\cr
&0.075 && 500 && 0.2 && 95.8 & \cr  
&0.1 && 1000 && 0.7 && 43.5 & \cr
&0.1 && 1500 && 1.0 && 17.5 & \cr  
&0.1 && 2000 && 1.4 && 9.4 & \cr 
&\omit&&\omit&&\omit&&\omit&\cr
\hline
\hline}}
\caption{Degenerate BESS at LHC for the process $p p\to L^\pm\to
\mu\nu_\mu+X$.}
\end{table}

\end{document}